\documentclass[a4paper,11pt]{article}
\usepackage{epsfig}

\begin{document}
\title{A general track reconstruction scheme and its application to the OPERA drift tubes}
\author{R. Zimmermann}
\maketitle
{\footnotesize Hamburg University, Physics Department, D-22761 Hamburg, Germany}\\[0.5cm]
\begin{center}
\end{center}
\begin{abstract}
A general reconstruction and calibration procedure for tracking and wire position determination of the OPERA drift tubes \cite{specbib} is presented. The mathematics of the pattern recognition and the track fit are explained.
\end{abstract}
\newpage
\section{Introduction}
The two most important parameters for a track reconstruction with drift tubes are the time--distance--relation ($rt$--relation) to convert the measured times into spatial distances, and the resolution function as a weight for the track fit. In this article the determination of these parameters is described for the OPERA drift tubes. 

OPERA is a long-baseline neutrino oscillation experiment \cite{operaprop} to search for the $\nu_{\mu} \rightarrow \nu_{\tau}$ oscillation in the parameter region indicated by previous experiments \cite{statop}. The main goal is to find $\nu_{\tau}$ appearance by direct detection of the $\tau$ from $\nu_{\tau}$ CC interactions. The OPERA detector consists of two massive lead-emulsion target sections followed by muon spectrometers. The task of the muon spectrometers, which consist of dipole magnets, RPCs\footnote{{\bf R}esistive {\bf P}late {\bf C}hamber} and $\sim$10000 aluminum drift tubes of 8\,m length \cite{specbib}, is to clarify the signature of the muonic $\tau$ decay  and to remove the background originating from charmed particles produced in $\mu$-neutrino interactions. 

It is a difficult task to obtain the parameters needed for the track reconstruction of the muons traversing the OPERA drift tubes
due to the very poor statistics of cosmic muons at the Gran Sasso laboratory where the OPERA detector is located. The determination of these parameters can, however, be done outside the underground lab using an equivalent detector with higher track statistics at same conditions (temperature, pressure, etc.). Since the mean cosmic energy at the sea level ($\approx 4$~GeV) is the same mean energy of the muons coming out from the tau lepton decay after the CC $\nu_{\tau}$ interaction in the target, a calibration setup was built and operated at Hamburg. For this two 8~m long drift modules (\cite{specbib}) with an total overlap in horizontal position were used, while the resulting wire sag was compensated by bending the tubes accordingly.
This article describes a calibration procedure to determine the $rt$--relation and the resolution function. Furthermore a wire alignment procedure is described.
\section{The pattern recognition}
\label{sec:pat}
\begin{figure}[b]
  \begin{center}
    \includegraphics[scale=0.45,clip]{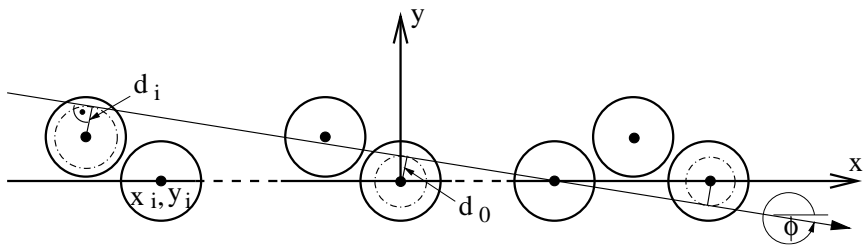}
  \end{center}
  \caption[Track description using the Hesse form.]{\label{picpattern}Track description using the Hesse form and definition of the parameters in the used coordinate system.}
\end{figure}
The pattern recognition fulfills two important functions. On the one hand it performs the preselection of hit candidates belonging to a track, on the other hand it delivers the start values for the track fit. 

We assume that a particle passing trough the detector (straight track) will produce $N$ hits with valid drift times $t$ within a defined time window. These times can be converted into distances $r_i(t)$ between the track and the fired wire by using the $rt$--relation. This radius $r_i(t)$ of a circle around the sense wire is an unsigned quantity and does not tell us on which side the particle passed the wire. The pattern recognition has to select the best sample of hit candidates, define the signs to resolve the right-left ambiguities and define the start angle and the start distance to the origin of the track. For the calibration this procedure has to be fast and efficient. Normally a Kalman filter\footnote{A Kalman filter is a stochastic state estimator for dynamic systems. The most probable start value will be predicted and afterward corrected with the measured one. The difference of both values will be linearly weighted and is used to improve the current state.} technique  \cite{kalbib} is used. But if a sufficient\footnote{The sufficient number depends on the system. For the OPERA drift tubes 5000 tracks are enough for the calibration.} number of events is available a more simple method described as follows fulfills the requirements above. It uses single straight tracks only. In case of more than one track per event (multi tracks) the event is ignored. 

For the track reconstruction with drift tubes the distance from the track to the wire is needed. Thus it is convenient to describe the particle track with the Hesse form
\begin{eqnarray}
  d_0 = x\,\sin \phi - y\,\cos \phi\, .\label{glhesse1}
\end{eqnarray}
Here $d_0$ is the track distance of the closest approach to the origin and $\phi$ the angle between track and x-axis as defined in Fig. \ref{picpattern}. In this analysis we assume parallel wires, therefore a two dimensional track description is sufficient. The distance of closest approach to the anode wires $d_i$ is then calculated by
\begin{eqnarray}
  d_i = d_0 - x_i\,\sin \phi + y_i\,\cos \phi.\label{glhesse2}
\end{eqnarray}
The index $i$ describes the wire number $i$ with the coordinates $x_i$ and $y_i$.

\begin{figure}[htbp]
  \begin{center}
    \includegraphics[scale=0.38,clip]{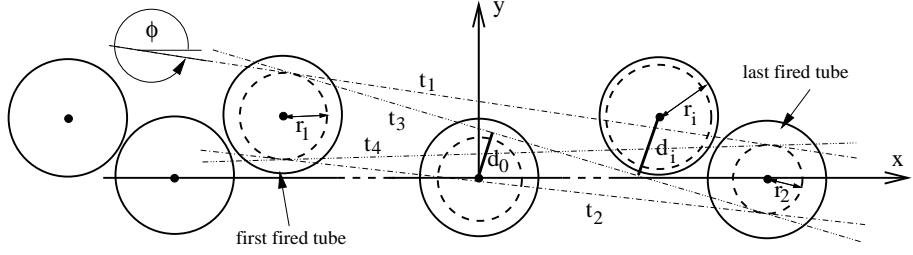}
  \end{center}
  \caption[Principle of pattern recognition.]{\label{picpat2}Example of the four tangents to the radii $r_1$ and $r_2$. The tangent, minimizing Eq. \ref{glchi1}, is the best description of the true track.}
\end{figure}
In the simplest pattern recognition scheme the two tubes with the maximum distance between each other in one event will be used (Fig. \ref{picpat2}). Now four possibilities exist to fit tangents ($t_1$ to $t_4$) to the radii $r_1$ and $r_2$. The tangent minimizing the $\chi^2$ expression will be selected
\begin{eqnarray}
  \chi^2 = \sum_{i=1}^N \frac{(r_i - d_i)^2}{\sigma^2}. \label{glchi1}
\end{eqnarray}
$\sigma$ is the mean resolution, which is assumed to be the same for all tubes in the pattern recognition. For starting $\sigma$ is set to 1.
\begin{figure}[htbp]
  \begin{center}
    \includegraphics[scale=0.4,clip]{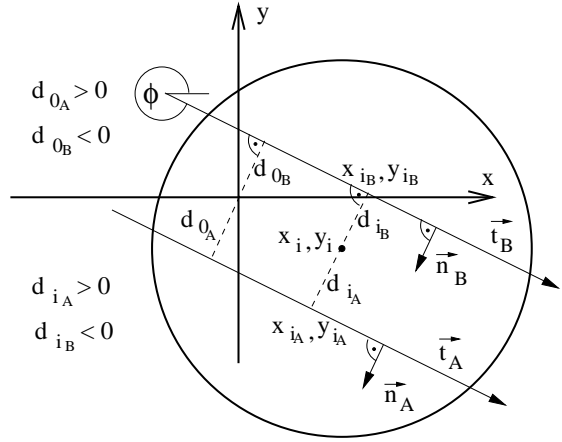}
  \end{center}
  \caption[Sign convention for the track description.]{\label{picpat3}Sign convention for the track description considering the cases {\em A} and {\em B} according to the particle track on the right or left side from the wire.}
\end{figure}
The minimal $\chi^2$ of the selected tangent has to be lower than a predefined maximal value (for this calibration values between 100 and 500 were used). For higher values the whole event will be rejected. The origin of such events can be noise or cross talk. If a valid tangent is found, the start parameter $d_{0_{Start}}$ and $\phi_{Start}$ are now available for the track fit. The signs of the $d_i$ will be defined using the convention shown in Fig. \ref{picpat3}. The normal vector $\vec{n}$ is defined as
\begin{eqnarray}
  \vec{n} = { \sin\phi \choose -\cos\phi}.
\end{eqnarray}
The particle can pass the wire at two sides, denoted by $A$ and $B$. Eq. \ref{glhesse1} can be written as
\begin{eqnarray}
  d_{0_{A,B}} = \vec{n} \hspace{2pt} \vec{x}_{i_{A,B}} = \sin \phi \hspace{2pt} x_{i_{A,B}} - \cos \phi \hspace{2pt} y_{i_{A,B}}. \label{glskalar1}
\end{eqnarray}
Here $x_{i_{A,B}}$ and $y_{i_{A,B}}$ are the points of contact to the tangents. The distances $d_i$ follow this convention, the vector $\vec{x}={x_i \choose y_i}$ in Eq. \ref{glskalar1} will be replaced by a vector pointing from the anode wire to the track
\begin{eqnarray}
  d_{i_{A,B}} = \vec{n} \hspace{2pt} (\vec{x}_i - \vec{x}_{i_{A,B}}). \label{glskalar2}
\end{eqnarray}

\section{Track fit}
\label{sec:trackFit}
\subsection{General track fitting}
\label{sec:gentrackFit}
In this section a general track fit procedure will be described. It was originally developed and used in the ARGUS experiment \cite{argusbib}. A former description for the outer tracker system of the HERA--B experiment can be found in \cite{bibzim}. 

The distances $d_{m,i}$ from the pattern recognition \footnote{The index $m$ denotes the distances calculated from the measured drift times.} form a vector $\vec{d}_m$ with the dimension $N$. The measurement uncertainties $\sigma_i$ for each used wire are collected in the resolution function, binned in time or space respectively. The squared values of this function are on the main diagonal of the $N \times N$ covariance matrix ${\bf V}$. In the case of independent measurements only the main diagonal is filled. The track will be described by the $N$ dimensional vector $\vec{d}_t(\vec{q})$. In general $\vec{q}$ is $M$ dimensional ($M$ is the number of track parameters, in our case two). The parameter $\vec{q}$ results from the fit of $\vec{d}_t(\vec{q})$ to $\vec{d}_m$. Using the least square method, the parameter $\vec{q}$ minimizes the expression
\begin{eqnarray}
\chi^2=\left[\vec{d}_m-\vec{d}_t(\vec{q})\right]^T {\bf W} \left[\vec{d}_m-\vec{d}_t(\vec{q})\right].\label{chi2a}
\end{eqnarray}
The weight matrix ${\bf W}$ is the inverse covariance matrix of the measured coordinates:
\begin{eqnarray*}
  {\bf W} = {\bf V}^{-1}.
\end{eqnarray*}
Assuming independent measurements equation \ref{chi2a} can be written as:
\begin{eqnarray}
  \chi^2=\sum_{i=1}^{N} \frac{1}{\sigma_i^2}(d_{m,i}-d_{t,i}(\vec{q}))^2.
\end{eqnarray}
Minimizing $\chi^2$ as a function of $\vec{q}$ is equivalent to finding a solution to the system of equations
\begin{eqnarray}
  \frac{\partial \chi^2}{\partial \vec{q}} = 0 .\label{minchi2}
\end{eqnarray}

In general these equations are non-linear because the track model $\vec{d}(\vec{q})$ is non-linear. However, in most of the cases it can be linearized and the equations can be solved iteratively. A linear track model is given by:
\begin{eqnarray}
  \vec{d}_t(\vec{q})={\bf A}\vec{q}+\vec{b} \hspace{1em} \mbox{with} \hspace{3em} {\bf A}=(A_{i\mu})_{i=1,...,N \atop{\mu = 1,...,M}}.\label{linchi}
\end{eqnarray}
Here $A$ denotes the Jakobi matrix. With this, Eq. \ref{chi2a} gives
\begin{eqnarray}
  \chi^2 & = & \left(\vec{d}_m-{\bf A}\vec{q}-\vec{b}\right)^T {\bf W} \left(\vec{d}_m-{\bf A}\vec{q}-\vec{b}\right) \\
         & = & \sum_{i=1}^{N} \frac{1}{\sigma_i^2}(d_{m,i}-\sum^{M}_{k=1} A_{i k}q_{k}-b_i)^2.\nonumber
\end{eqnarray}
By differentiation one gets
\begin{eqnarray}
\frac{\partial \chi^2}{\partial q_k} & = & -2\sum_{i=1}^{N} A_{i\mu}\frac{1}{\sigma_i^2}(d_{m,i}-\sum^{M}_{k=1} A_{ik}q_k-b_i) \nonumber \\
& = & -2\left[{\bf A}^T {\bf W} \left(\vec{d}_{m}-{\bf A}\vec{q}-\vec{b}\right)\right]_{\mu} \\
  \Rightarrow \frac{\partial \chi^2}{\partial \vec{q}} & = & -2{\bf A}^T {\bf W} \left(\vec{d}_{m}-{\bf A}\vec{q}-\vec{b}\right) \stackrel{Def}{=} 0.
\end{eqnarray}
This yields the parameters
\begin{eqnarray}
  \fbox{$\vec{q} = \left({\bf A}^T {\bf W} {\bf A}\right)^{-1} {\bf A}^T {\bf W} \left(\vec{d}_{m}-\vec{b}\right).$}
\end{eqnarray}
With the found parameters one can solve the following equation system:
\begin{eqnarray}
  {\bf A}^T {\bf W} {\bf A} \vec{q} =  {\bf A}^T {\bf W} \left(\vec{d}_{m}-\vec{b}\right).\label{Glsys}
\end{eqnarray}
From the general formula for error propagation (\cite{BLO83})
\begin{eqnarray}
  {\bf V}(\vec{y}) = {\bf ZV}(\vec{x}){\bf Z}^T \hspace{2em} \mbox{for} \hspace{3em} \vec{y}={\bf Z}\vec{x} + \vec{c}
\end{eqnarray}
and the replacements
\begin{eqnarray*}
  \vec{x} & \equiv & \vec{d}_{m} \\
  {\bf V}(\vec{x}) & \equiv & {\bf W}^{-1} \\
  \vec{y} & \equiv & \vec{q} \\
  {\bf Z} & \equiv & ({\bf A}^T{\bf WA})^{-1}{\bf A}^T{\bf W}
\end{eqnarray*}
one receives the covariance matrix
\begin{eqnarray}
  \fbox{${\bf V}(\vec{q}) = ({\bf A}^T{\bf WA})^{-1}.$}
\end{eqnarray}
As already mentioned this method can be used for a non-linear track model if it can be linearized. A linearisation requires a starting value $\vec{q}_0$ near the true solution. In this case one can describe the vector $\vec{q}_0$, which essentially contains the non-linearities, with the first order of the Taylor expansion
\begin{eqnarray}
  \left. \vec{d}_t(\vec{q}) \approx \vec{d}_t(\vec{q_0}) + \sum_{\mu =1}^M \frac{\partial \vec{d}_t}{\partial q_{\mu}} \right|_{\vec{q}=\vec{q_o}} (q_{\mu}-q_{0,\mu}) = \vec{d}_t(\vec{q_0}) + {\bf A} (\vec{q}-\vec{q_0}).
\end{eqnarray}
The Jakobi matrix of $\vec{d}_t(\vec{q})$ for $\vec{q}=\vec{q_0}$ is defined as
\begin{eqnarray}
  \left. {\bf A} = \frac{\partial \vec{d}_t}{\partial \vec{q}} \right|_{\vec{q}=\vec{q_o}} = (A_{i\mu}) = \left( \frac{\partial d_{t_i}}{\partial q_{\mu}}\right)_{i=1,...,N \atop{\mu = 1,...,M}}.\label{gljakobim}
\end{eqnarray}
Correspondingly \hspace{1em} $\vec{b} \equiv \vec{d}_t(\vec{q}_0) - {\bf A} \vec{q}_0$ \hspace{1em} (see Eq. \ref{linchi}) can be used in Eq. \ref{Glsys}. This results
\begin{eqnarray}
  {\bf A}^T {\bf W} {\bf A} (\vec{q}-\vec{q}_0) = {\bf A}^T {\bf W} \left[\vec{d}_{m}-\vec{d}_t(\vec{q}_0)\right].\label{glsys2}
\end{eqnarray}
The change of parameters
\begin{eqnarray}
  \Delta \vec{q} \equiv \vec{q} - \vec{q}_0
\end{eqnarray}
results into a better approximation for the next iteration
\begin{eqnarray}
  \vec{q}_1 = \vec{q}_0 + \Delta \vec{q}.\label{parch}
\end{eqnarray}
Eq. \ref{glsys2} and \ref{parch} define an iterative procedure. In the $n$th iteration, equation
\begin{eqnarray}
  \fbox{${\bf A}^T {\bf W} {\bf A} (\Delta \vec{q}) = {\bf A}^T {\bf W} \left[\vec{d}_{m}-\vec{d}_t(\vec{q}_{n-1})\right].$}\label{glsys3}
\end{eqnarray}
is solved for $\Delta \vec{q}$. The vector $\vec{d}_m(\vec{q}_{n-1})$ contains the fitted track distances for the track parameters of the next iteration and ${\bf A}$ is the Jakobi matrix of the track parameter for $\vec{q} = \vec{q}_{n-1}$. 
$\Delta \vec{q}$ then gives the new approximation of the parameters
\begin{eqnarray}
  \vec{q}_n = \vec{q}_{n-1} + \Delta \vec{q}.
\end{eqnarray}
The end condition is reached when $\Delta \vec{q} \approx 0$ and
\begin{eqnarray}
  \chi^2_n \approx \chi^2_{n-1}.
\end{eqnarray}

\subsection{Track fit with two parameters}\label{sec:fit}
In this section a minimal track fit with two parameters will be described in detail for the drift tube modules. The initial parameter $\vec{q}_{Start}$ for the track fit found by the pattern recognition is 
\begin{eqnarray}
  \vec{q}_0 = \vec{q}_{Start} = {d_{0_{Start}} \choose \phi_{Start}}. 
\end{eqnarray}
The Jakobi matrix (Eq. \ref{gljakobim}) can be written as:
\begin{eqnarray}
  \left. {\bf A} = \frac{\partial \vec{d}}{\partial \vec{q}} \right|_{\vec{q}=\vec{q}_{n-1}} = (A_{i\mu}) = \left( \frac{\partial d_i}{\partial q_{\mu}}\right)_{i=1,...,N \atop{\mu = 1,2}}. \label{gljakobi}
\end{eqnarray}
$(A_{i\mu})$ contains the partial derivatives of $d_i$ (see Eq. \ref{glhesse2}) with respect to $d_0$ ($\mu = 1$) and $\phi$ ($\mu = 2$)
\begin{eqnarray}
  \frac{\partial d_i}{\partial d_0} = 1 \hspace{2em} \mbox{and} \hspace{2em} \frac{\partial d_i}{\partial \phi} = -x_i \hspace{2pt} \cos \phi - y_i \hspace{2pt} \sin \phi.
\end{eqnarray}
By substitution (Eq. \ref{glsys3})
\begin{eqnarray}
  {\bf {\cal G}} = {\bf A}^T {\bf W A} \hspace{2em} \mbox{and} \hspace{2em} {\bf {\cal Y}} = {\bf A}^T {\bf W} (\vec{d}_{m} - \vec{d}_t(\vec{q}_{n-1})), \label{glfit1}
\end{eqnarray}
one receives
\begin{eqnarray}
  {\bf {\cal G}}\Delta \vec{q} = {\bf {\cal Y}} \hspace{2em} \mbox{and} \hspace{2em} \Delta \vec{q} = {\bf {\cal G}^{-1}{\cal Y}}. \label{glfit2}
\end{eqnarray}
${\bf {\cal G}}$ is an $2 \times 2$ matrix while ${\bf {\cal Y}}$ and $\vec{q}$ are two dimensional vectors. For independent measurements ${\bf {\cal G}}$ and ${\bf {\cal Y}}$ can be calculated by
\begin{eqnarray}
  \vspace*{2em}
  {\bf {\cal G}}_{\mu \nu} & = & \displaystyle \sum\limits_{i=1}^{N} \frac{1}{\sigma_i^2} \frac{\partial d_i}{\partial q_{\mu}} \frac{\partial d_i}{\partial q_{\nu}} \hspace{2em} \mu,\nu = 1,2 ,\label{glgy1} \\[1em]
  {\bf {\cal Y}}_{\mu} & = & \displaystyle\sum\limits_{i=1}^{N} \frac{\partial d_i}{\partial q_{\mu}} \frac{1}{\sigma_i^2} (d_{m,i} - d_{t,i}).\label{glgy2}
  \vspace{2em}
\end{eqnarray}
The calculation of the partial derivatives is done in Eq. \ref{gljakobi}. The subscripts $\mu$ and $\nu$ describe the elements of the parameter vector $\vec{q}$. Calculating $\Delta \vec{q}$ with Eq. \ref{glfit2} gives the new iterated parameters
\begin{eqnarray}
\vec{q}_n = \vec{q}_{n-1} + \Delta \vec{q}.
\end{eqnarray}
This procedure (Eq. \ref{glfit2}) will be repeated, until the end condition
\begin{eqnarray}
  \Delta \chi^2 = \chi^2_{n-1} - \chi^2_n < \Delta \chi^2_{min} \label{gldchi2}
\end{eqnarray}
is fulfilled.
\begin{figure}[htbp]
  \begin{center}
    \includegraphics[scale=0.6,clip]{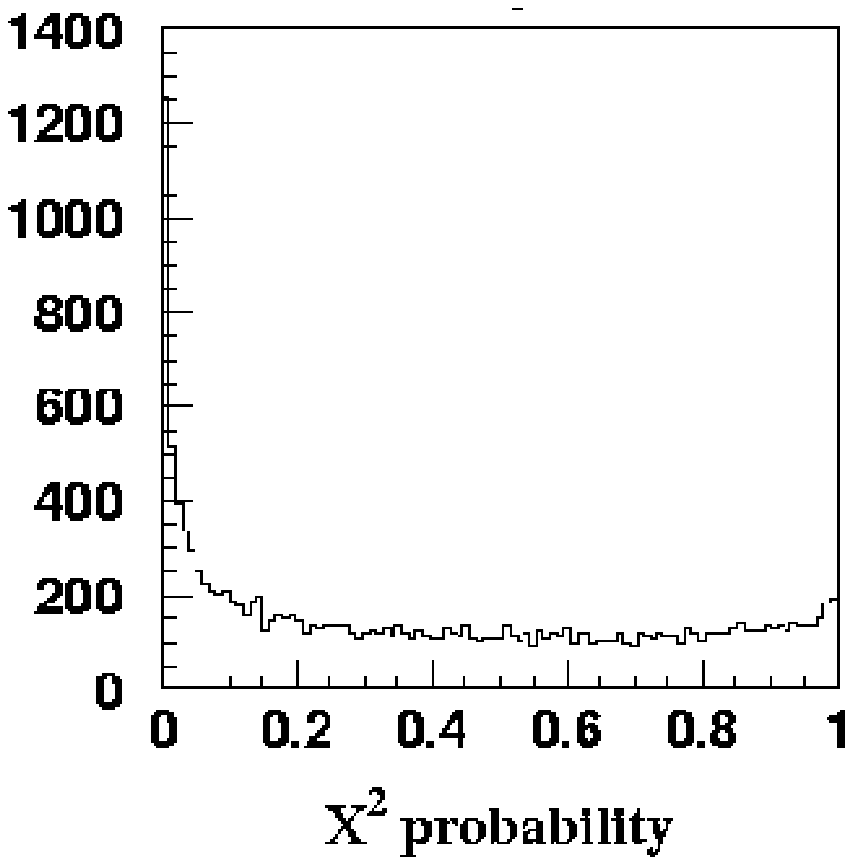}
    \includegraphics[scale=0.16,clip]{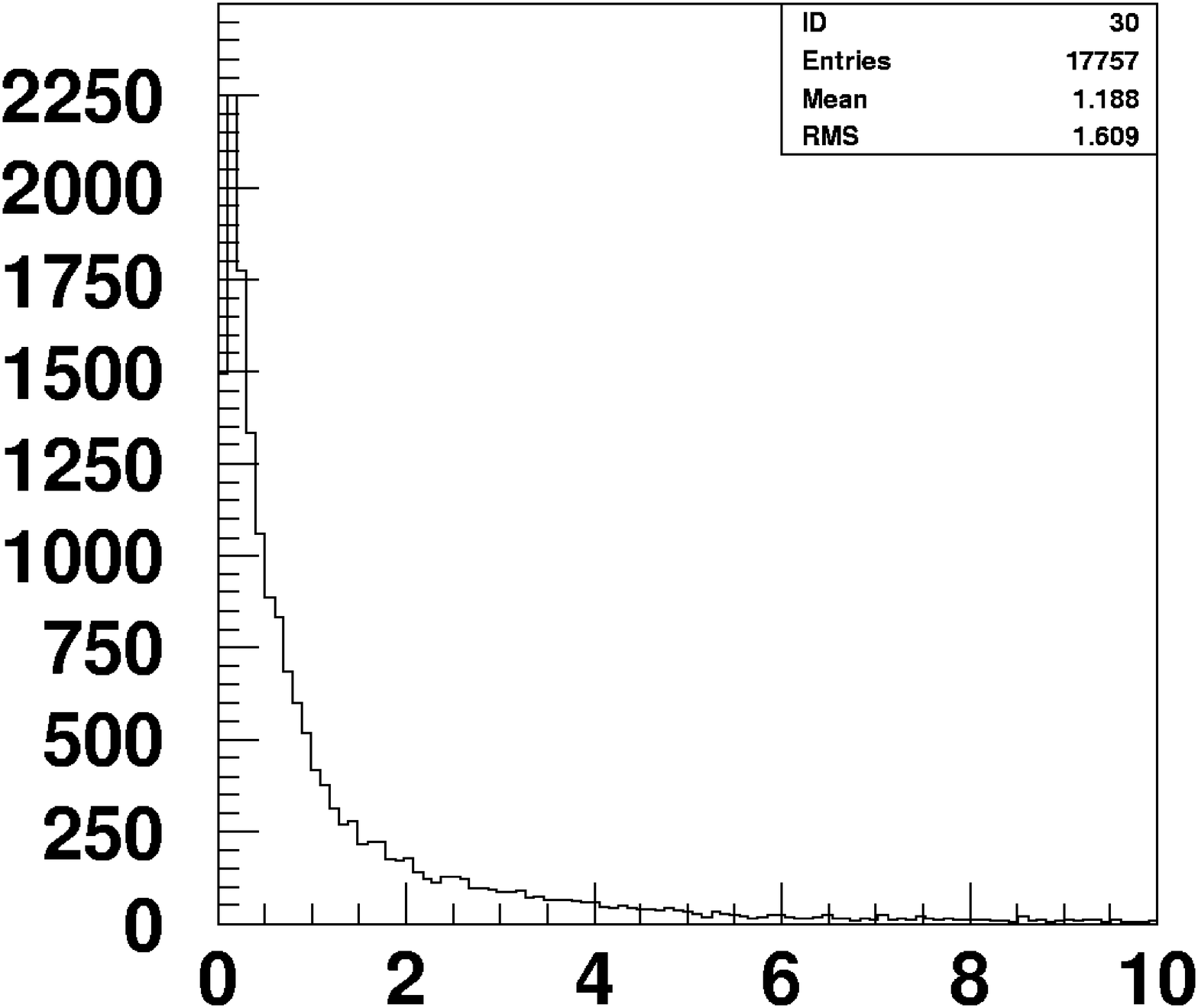}
  \end{center}
  \caption[Probability distribution of $\chi^2$]{\label{picprob}Probability distribution of $\chi^2$ (left) and the distribution of $\chi^2$ per degree of freedom (right).}
\end{figure}
A quality check for the track fit is the distribution of the $\chi^2$ probability, shown in Fig. \ref{picprob}. On the left side events have been clustered with an under--estimated resolution whereas the resolution has been over--estimated for the events on the right side. Except for the peak on the left side due to noise and cross talk, this distribution is flat, since a uniform error distribution is expected. Furthermore, the mean value of the $\chi^2$ distribution per degree of freedom, also shown in Fig. \ref{picprob}, should be around one if the track fit works correctly. In reality the mean value varies between 1 and 1.3.

For the fit described here a $\Delta \chi^2_{min} = 10^{-7}$ was used. The procedure converges usually after three iterations. After that the signs of $d_{m,i}$, determined by the pattern recognition, will be corrected (turned) if they are opposite to the signs of $d_{t_i}$. This procedure is repeated until no sign changes appear anymore (solution of left/right ambiguity). Such hits with random time signals, coming from noise or cross talk, are removed if the $\chi^2$ distribution is too large. The whole procedure is repeated, starting with the $\vec{q}_0$ from the pattern recognition, until $\chi^2 \le \chi^2_{max}$. After this it will be tested if still enough tubes $N$ are left over for the reconstruction ($>N_{tubes_{min}}=4$). In this case, the vector $\vec{q}_n$ of the last iteration contains the track parameter with the best track description. If not, the whole event is rejected.
\section[Calibration]{Calibration}\label{seckal}
The calibration is the iterative determination of parameters like the $rt$ relation and the resolution function $\sigma(t_D)$. In addition the alignment of the wire positions is possible with this procedure and will be described. 

The calibration procedure is based on the concept of residuals $\epsilon (t_D)$, defined as difference of the fitted drift distance ($d_t$) and the measured one ($d_m$)
\begin{eqnarray}
  \epsilon(t_D) \equiv d_t - d_m. \label{glresid}
\end{eqnarray}
The sign of $d_m$ is defined as sign of $d_t$.
\begin{figure}[htbp]
  \begin{center}
    \includegraphics[scale=0.4,clip]{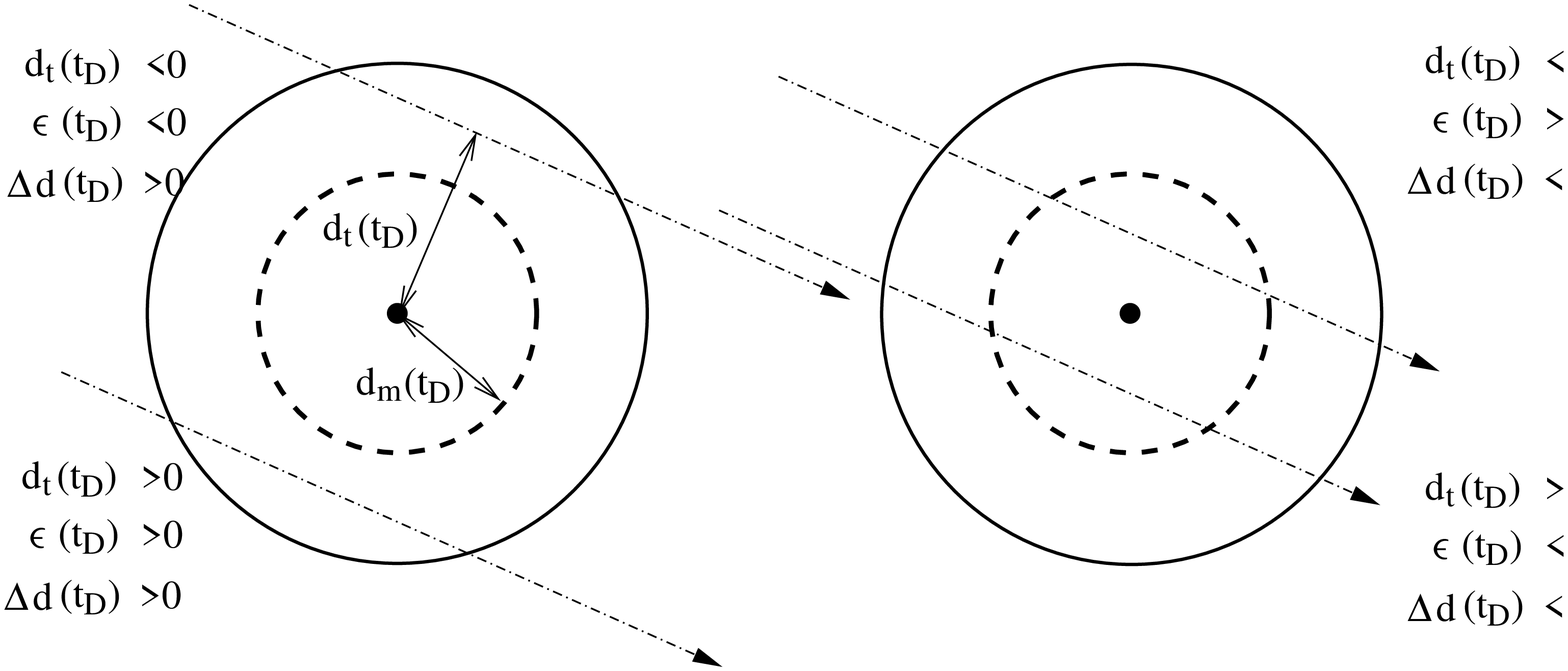}
  \end{center}
  \caption[The signs of the drift distances, the residuals and the corrections of the $rt$--relation.]{\label{picvorz} The signs of the drift distances, the residuals and the corrections of the $rt$--relation.}
\end{figure}
The $rt$--relation and the resolution function are represented by subdividing the parameters in time bins for an iterative procedure.
\subsection[Time distance relation]{Time distance relation}\label{sssecdob}
The correlation of the drift time $t_D$ and the drift distance is given by the $rt$--relation using the drift velocity $v_D$
\begin{eqnarray}
  d_{t}(t_D)=\int_0^{t_D}v_D(t) \,dt = \int_0^{t_D} \frac{dr}{dt} \,dt .\label{gldob}
\end{eqnarray}
The iterative correction of the $rt$--relation is done by the difference $\Delta d_m(t_D)$, calculated from the residuals including the signs of the drift distances
\begin{eqnarray}
  \Delta d_m(t_D) \equiv sign(d_t) \, \epsilon(t_D) = sign(d_t)(d_t - d_m) . \label{gldeldob}
\end{eqnarray}
$\Delta d_m(t_D)$ specifies, whether the measured drift distance is too low or too high. The signs used in the above equations are presented in Fig. \ref{picvorz}. The drift distance $d_t(t_D)$ is defined negative, if the particle track passed the wire on the left side (according to the defined coordinates) and positive on the right side. The sign of the residuals, arising from the drift distances, follows the same convention if the distance of the fitted track is larger than the measured drift distance. In the other case, the sign of the residuals is opposite to the drift distances. If $\Delta d_m(t_D) > 0$, the $rt$--relation is increased in the given time bin for the next iteration, for $\Delta d_m(t_D) < 0$ it is decreased. Normally the residuals are Gaussian distributed. In the vicinity of the wire the influence of the primary ionisation statistics increases due to the cluster distribution, and the drift time distribution will be asymmetric to larger values and thus to larger drift distances. Using Eq. \ref{gldeldob}, the distribution of $\Delta d_m(t_D)$ will be asymmetric to the negative side. This effect will be stronger the closer the track passes the wire. Since the difference in Eq. \ref{glresid} can not be smaller than the negative value of $d_m$ (same sign as $d_t$), this effect leads to a cut on the negative side in the $\Delta d_m(t_D)$ distribution (see chapter \ref{sec:fit}).
\begin{figure}[htbp]
  \begin{center}
      \includegraphics[scale=0.6,clip]{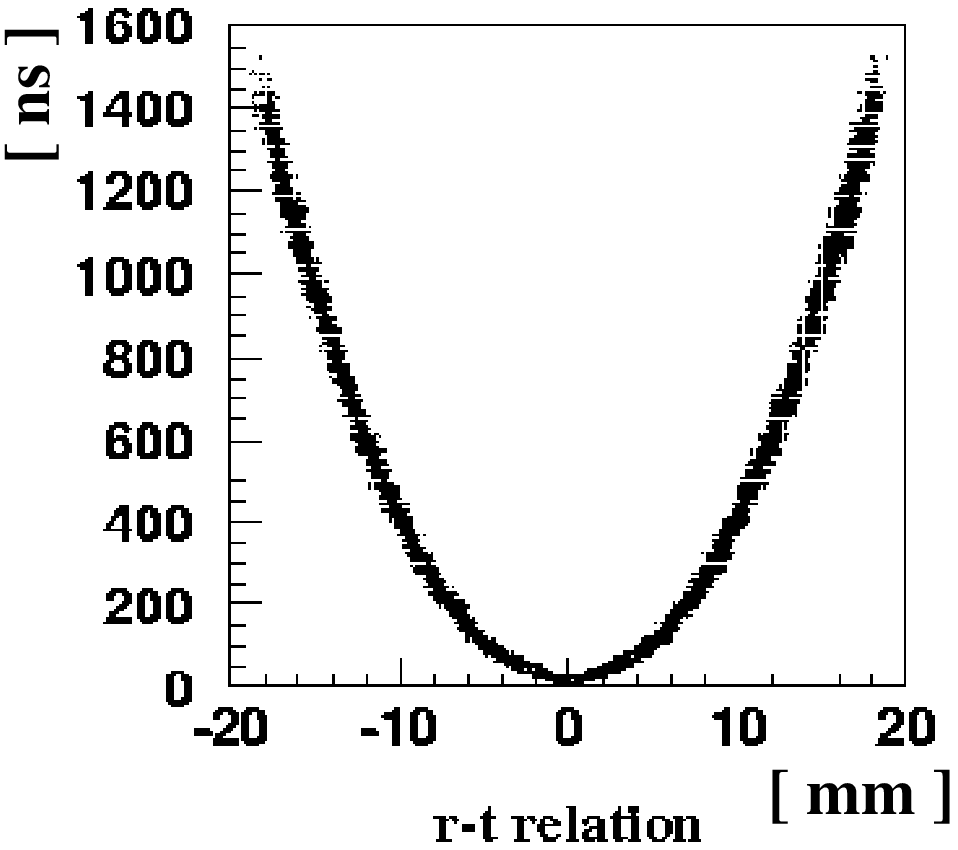}
      \includegraphics[scale=0.3,clip]{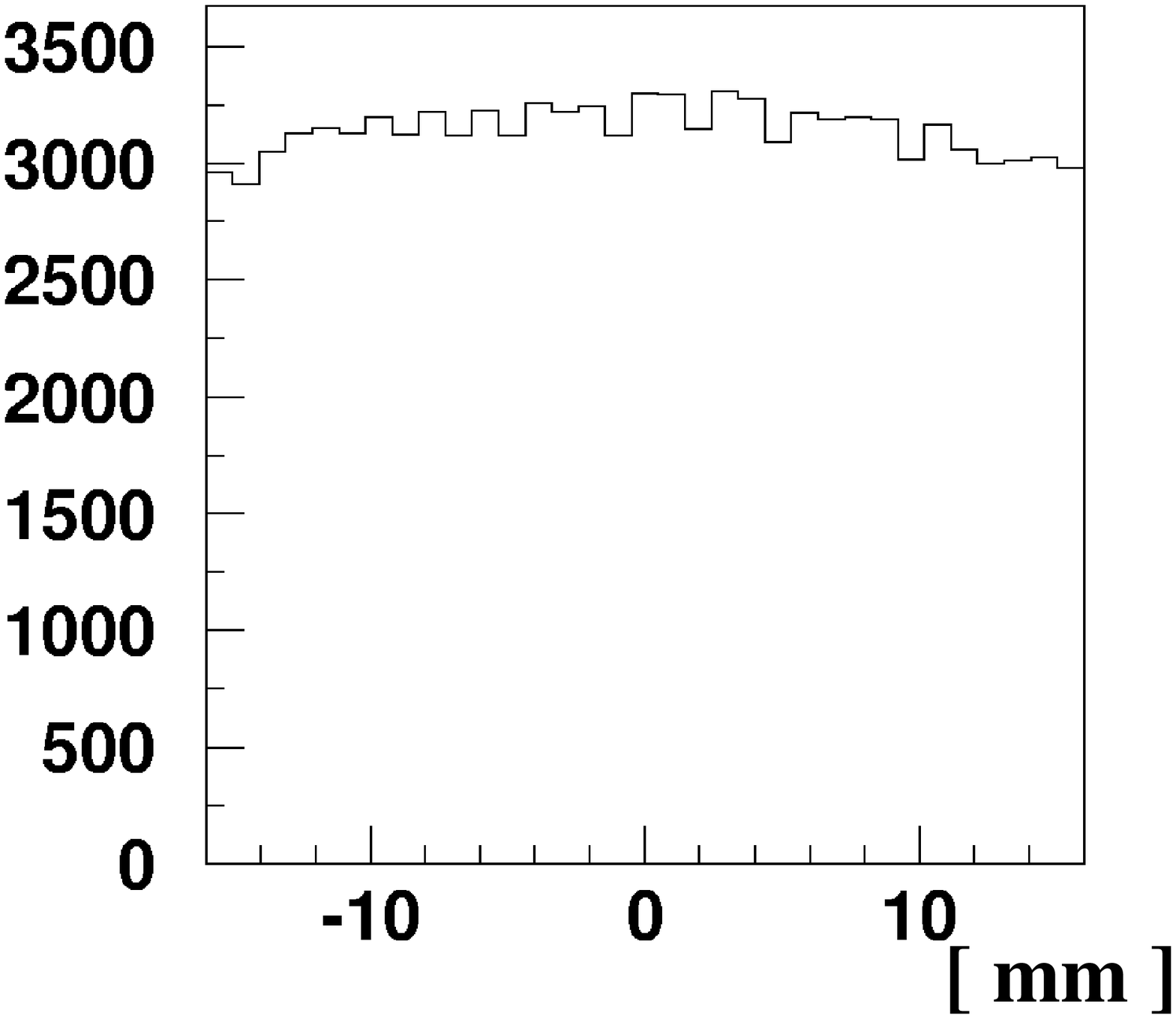}
  \end{center}
  \caption[$rt$--relation]{Correlation between drift times and distances, called $rt$--relation (left). On the right side the distribution of the fitted distances are shown.\label{picdob2}}
\end{figure}
The cut on $\Delta d_m(t_D)$ distribution is exactly $-|d_m|$. Thus the distribution is positive at the wire position. In all cases the maximal entries are at zero position.
The mean value of $\Delta d_m(t_D)$ corrects the $rt$--relation as follows
\begin{eqnarray}
  d_m(t_D)_{new} = d_m(t_D)_{old} + \langle\!\hspace{2pt}\Delta d_m(t_D)\!\hspace{2pt}\rangle, \label{gldobcor}
\end{eqnarray}
with $\langle\!\hspace{2pt}\Delta d_m(t_D)\!\hspace{2pt}\rangle$ as mean value of the $\Delta d_m(t_D)$ distribution. The iteration stops if the squared sum over all $\Delta d_m(t_D)$ is not changing with further iterations.

For the start, the distance $d_m(t_D)_{old}$ is obtained by integration of the drift time spectrum. In a first approximation a drift tube with a homogeneous distribution of track distances will be assumed.
With $dN$ tracks, passing the wire within the interval $[r;r+dr]$,
\begin{eqnarray}
  \frac{dN}{dr} = const = \frac{N_{track}}{r_{tube}} ,\label{glspurdens}
\end{eqnarray}
$N_{track}$ is the number of tracks which hit the tube and $r_{tube}$ is the tube radius. The drift velocity $v_D$ is given by
\begin{eqnarray}
  v_D = \frac{r_{tube}}{N_{track}} \cdot \frac{dN}{dt}.
\end{eqnarray}
By integration of the drift time spectrum $dN/dt$ and normalizing to the maximal tube radius one gets a first estimation of the $rt$--relation
\begin{eqnarray}
  d_{m,start}(t_D)=\frac{r_{tube}}{N_{track}} \int_0^{t_D} \frac{dN}{dt} \,dt. \label{gldobnah}
\end{eqnarray}

Fig. \ref{picdob2} shows the correlations between the distances $d_t(t_D)$ and the drift times $t_D$. The mean values of the $d_t(t_D)$ distributions of both branches form the $rt$--relation. For the calibration two versions of $rt$--relations are used. The description of the time binned version is a good approach for distances not too close to the wire. Since the residua distribution near the wire is asymmetric as mentioned above, the $rt$--relation (time binned) is bend away from the actual wire position. This description is unrealistic, since the position zero is existing. To avoid this behavior, a combination of time binned and space binned $rt$--relation is used, where the space binned relation is used for positions near the wire. Thus the wire position is reached, but the drift times are always greater than zero. This behavior is realistic, since a drift time zero can not appear due to the cluster distribution even in the case if the wire is hit by a track. For describing, interpolating and smoothing the $rt$--relation cubic splines are used. If the method is working correctly, the distribution of the fitted distances is theoretically flat. In reality this distribution is a bit bended, but should be without spikes and gaps.
\subsection[Spatial resolution]{Spatial resolution}\label{sssecaufl}
For the determination of the resolution function $\sigma_j (t_D)$ the RMS value of the residual distribution $\epsilon_i(t_D)$ will be used
\begin{eqnarray}
  \sigma_j (t_D) = \sqrt{\frac{1}{N} \sum_{i=1}^N (f_i \epsilon_i(t_D))^2}.\label{glresol}
\end{eqnarray}
$N$ specifies the entries of each residual distribution for a given time bin $j$ and $f_i$ is the residual scale factor.
Since the residuals are calculated as the difference of measured and fitted track distances, they contain the uncertainties of the fitted track. To avoid underestimating the resolution\footnote{Only the fitted track is known, not the real one.} a scale factor for correction is needed. The calculation of this factor will be described in the following \cite{KAP94}.

\begin{figure}[htbp]
  \begin{center}
      \includegraphics[scale=0.75,clip]{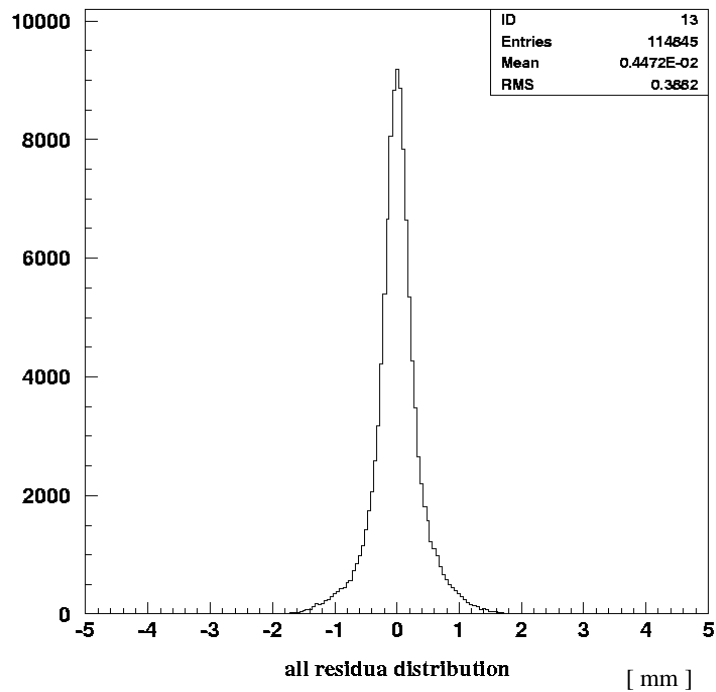}
  \end{center}
  \caption[Distribution of the residuals corrected by the scale factor. The RMS value yields the mean spatial resolution.]{\label{picfac}Distribution of the residuals corrected by the scale factor. The RMS value yields the mean spatial resolution.}
\end{figure}

The resolution function is used to construct the covariance matrix ${\bf V}(\vec{d}_m)$ which appears in the $\chi^2$ expression (Eq. \ref{chi2a}). The expectation of the squared deviations of measured and fitted distances to the wires are the diagonal matrix elements
\begin{eqnarray}
  V_{ii}(\vec{d}_m) = \sigma^2 [d_{m,i}(t_D)].
\end{eqnarray}
Since per definition the true track (parameter $\vec{q}_{true}$) has no errors, the covariance matrix can be written as
\begin{eqnarray}
  {\bf V}(\vec{d}_m) = {\bf V}(\vec{d}_m - \vec{d}_t(\vec{q}_{true})) = {\bf V}(\vec{\epsilon}_{true}).
\end{eqnarray}
That means the covariance matrix elements are equal to the standard deviation of the true residual distribution. Since the true track (and the true residual distribution) is unknown, it will be replaced by the parameter $\vec{q}$ of the fitted track of the last iteration. The relation of the measured residuals to the covariance matrix is
\begin{eqnarray}
  {\bf V}(\vec{\epsilon}) = {\bf V}(\vec{d}_m - \vec{d}_t(\vec{q})) \ne {\bf V}(\vec{d}).
\end{eqnarray}
By linearisation
\begin{eqnarray}
  d(\vec{q}) = d(\vec{q}') + {\bf A}(\vec{q} - \vec{q}')
\end{eqnarray}
the true residuals can be expressed by measurable values
\begin{eqnarray}
  \vec{\epsilon}_{true} = \vec{d}_m\! -\!\vec{d}_t(\vec{q}_{true})\!=\! \vec{d}_m\! -\! \left[\vec{d}_t(\vec{q})\! +\! {\bf A}(\vec{q}_{true}\! -\! \vec{q})\right]\! = \vec{\epsilon} -\! {\bf A}(\vec{q}_{true}- \vec{q}).
\end{eqnarray}
By means of a general error propagation (\cite{BLO83}) one gets
\begin{eqnarray}
  {\bf V}(\vec{\epsilon}_{true}) = {\bf V}(\vec{\epsilon}) + {\bf V}({\bf A}\vec{q}) = {\bf V}(\vec{\epsilon}) + {\bf AV}(\vec{q}){\bf A}^T .
\end{eqnarray}
This means, the reference trajectory $\vec{d}_t(\vec{q})$ is only known within the track parameter errors. If only the standard deviation of the measured residual distribution is used for the correction of the resolution function, it will be underestimated (further information in \cite{EIC81}). A convenient correction is done by the residual scale factor. With the requirement
\begin{eqnarray}
  V_{ii}(\vec{\epsilon}_{true}) \stackrel{Def}{=} f^2_i \cdot V_{ii}(\vec{\epsilon}) = f^2_i \cdot \left [ V_{ii}(\vec{\epsilon}_{true}) - ({\bf AV}(\vec{q}) {\bf A}^T)_{ii} \right]
\end{eqnarray}
one gets the factor
\begin{eqnarray}
  \fbox{$ f_i = \displaystyle \sqrt{\frac{\sigma^2_i}{\sigma^2_i - ({\bf AV}(\vec{q}) {\bf A}^T)_{ii}}}  $}\label{glfac}
\end{eqnarray}
for the $i$th measured residual $\epsilon_i$ of the track. ${\bf A}$ is the Jakobi matrix (Eq. \ref{gljakobi}) and ${\bf V}={\bf W}^{-1}$ the covariance matrix of the track parameter vector $\vec{q}$ (Eq. \ref{glfac}). The field $\sigma_i$ contains the calculated resolution (Eq. \ref{glresol}) for the given drift time bin of the last iteration. Fig. \ref{picfac} shows the residual distribution, corrected with the scale factor. If the complete covariance matrix is unknown, a global correction factor can be used. Assuming that all $N$ fired tubes have the same Gaussian distributed error $\sigma$ and the same mean squared measured residual $\epsilon^2$, one gets
\begin{eqnarray}
  \chi^2_{min} = N \frac{\epsilon^2}{\sigma^2} = n_{dof} = N-2
\end{eqnarray}
with $n_{dof}$ degrees of freedom (2 for a linear track model). 

\begin{figure}[htbp]
  \begin{center}
      \includegraphics[scale=0.78,clip]{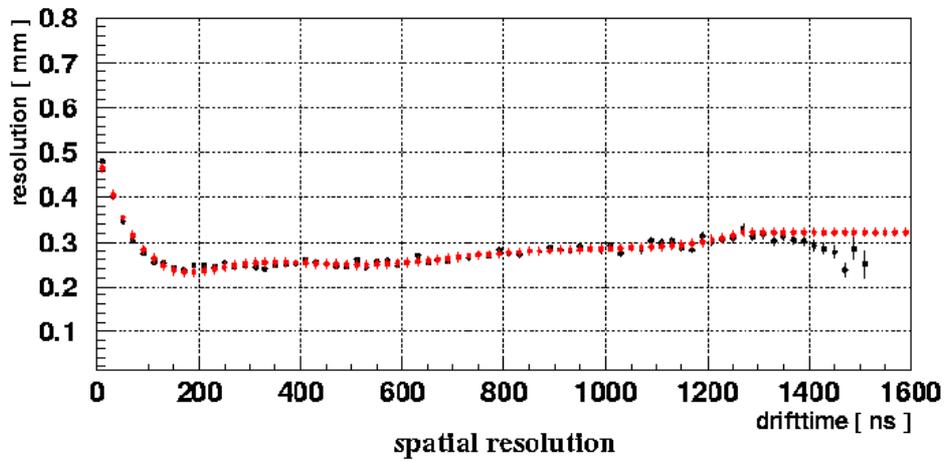}
  \end{center}
  \caption[Resolution function dependent on the drift time.]{\label{picaufl}Resolution function (squares) dependent on the drift time, fitted with cubic splines (diamonds).}
\end{figure}

Each residual $\epsilon_i(t_D)$ will be corrected by the factor and filled into histograms. The RMS value of this distribution is the mean resolution $\sigma (t_D)$.
Repetition for each time bin $i$ yields the resolution function $\sigma_i(t_D)$ dependent on the drift time. The weight requires a resolution for each individual wire distance $d$ and $d_t$. For the interpolation and smoothing cubic splines are used. Thus resolution values for a finer binning are available. Fig. \ref{picaufl} shows the resolution function and the splines.
The time dependent resolution function can be converted into a distance dependent one by using the $rt$--relation.

A quality factor of drift tubes is the mean resolution
\begin{eqnarray}
  \langle \sigma \rangle = \frac{1}{\sqrt{\displaystyle \frac{1}{N} \sum_{i=1}^N \frac{1}{\sigma^2_i}}},\label{glavaufl}
\end{eqnarray}
with $\sigma_i$ as resolution of the $i$th of $N$ drift time bins (e.g.  20~ns per bin). This definition takes into account, that distances with a better resolution have a higher weight in the track fit. 
Using the mean resolution $\langle \sigma \rangle$ in the next iteration of the track fit, the mean errors of the track parameters are equal to the weighting of a variable resolution function $\sigma(t_D)$ \cite{KAP94}.

The resulting resolution function can be described by a theoretical model 
\begin{equation}
\sigma(d)_{theo} = \sqrt{\frac{j^3}{2n_P^2(2n_P^2r^2+j^2)} + k^2 + d^2} . \label{glauflmod}
\end{equation}
It consists of contributions from the primary ionisation statistics, from the diffusion and from a constant part $k$ (uncertainties in wire position, time jitter in the electronics, time resolution, etc.). The diffusion, dependent on the field and the drift distance, is specified by $d$. The parameter $j$ is the number of the triggered cluster, $n_P$ the number of primary ionisations per unit and $r$ the drift distance.
\subsection[Wire calibration]{Wire calibration}
The presented procedure can also be used for the calibration of the wire positions by replacing the track parameter vector $\vec{q}$ with a vector $\vec{\beta}$ describing the wire positions 
\begin{eqnarray}
  \vec{\beta} = {\beta_x \choose \beta_y}.
\end{eqnarray}
The track description is then given by
\begin{eqnarray}
  d_i = d_0 - \beta_x\,\sin(\phi) + \beta_y\,\cos(\phi).\label{glhesse2w}
\end{eqnarray}
The Jakobi matrix can be written as
\begin{eqnarray}
  \left. {\bf A} = \frac{\partial \vec{d}}{\partial \vec{\beta}} \right|_{\vec{\beta}=\vec{\beta}_{n-1}} = (A_{i\mu}) = \left( \frac{\partial d_i}{\partial \beta_{\mu}}\right)_{i=1,...,N \atop{\mu = 1,2}}, \label{gljakobiw}
\end{eqnarray}
with
\begin{eqnarray}
\frac{\partial d_i}{\partial \beta_{x,j}} = -\delta_{ij} \sin \phi \hspace{2em} \mbox{and} \hspace{2em} \frac{\partial d_i}{\partial \beta_{y,j}} = \delta_{ij} \cos \phi.
\end{eqnarray}
$i$ and $j$ denote the wire number. Equation \ref{glsys3} can be written as
\begin{eqnarray}
  {\bf A}^T {\bf W} {\bf A} (\Delta \vec{\beta}) = {\bf A}^T {\bf W} \left[\vec{d}_{m}-\vec{d_t}(\vec{\beta}_{n-1})\right].\label{gliterpw}
\end{eqnarray}
$\Delta \vec{\beta}$ contains the wire shifting in the $x$ and $y$ direction. The calculation of $\Delta \vec{\beta}$ is done by
\begin{eqnarray}
  {\bf {\cal G}}\Delta \vec{\beta} = {\bf {\cal Y}} \hspace{2em} \mbox{and} \hspace{2em} \Delta \vec{\beta} = {\bf {\cal G}^{-1}{\cal Y}}, \label{glfit2w}
\end{eqnarray}
with the determination of ${\bf {\cal G}}$ and ${\bf {\cal Y}}$ as in  Equations \ref{glfit1}, \ref{glgy1} and \ref{glgy2} by replacing $\vec{q}$ with $\vec{\beta}$. The stop condition is the same as described by Eq. \ref{gldchi2}. In this case $\Delta \beta$ should not change (after 2 - 3 iterations). For a start parameter the theoretical wire positions are used.

\begin{figure}[htbp]
  \begin{center}
    \includegraphics[scale=0.23,clip]{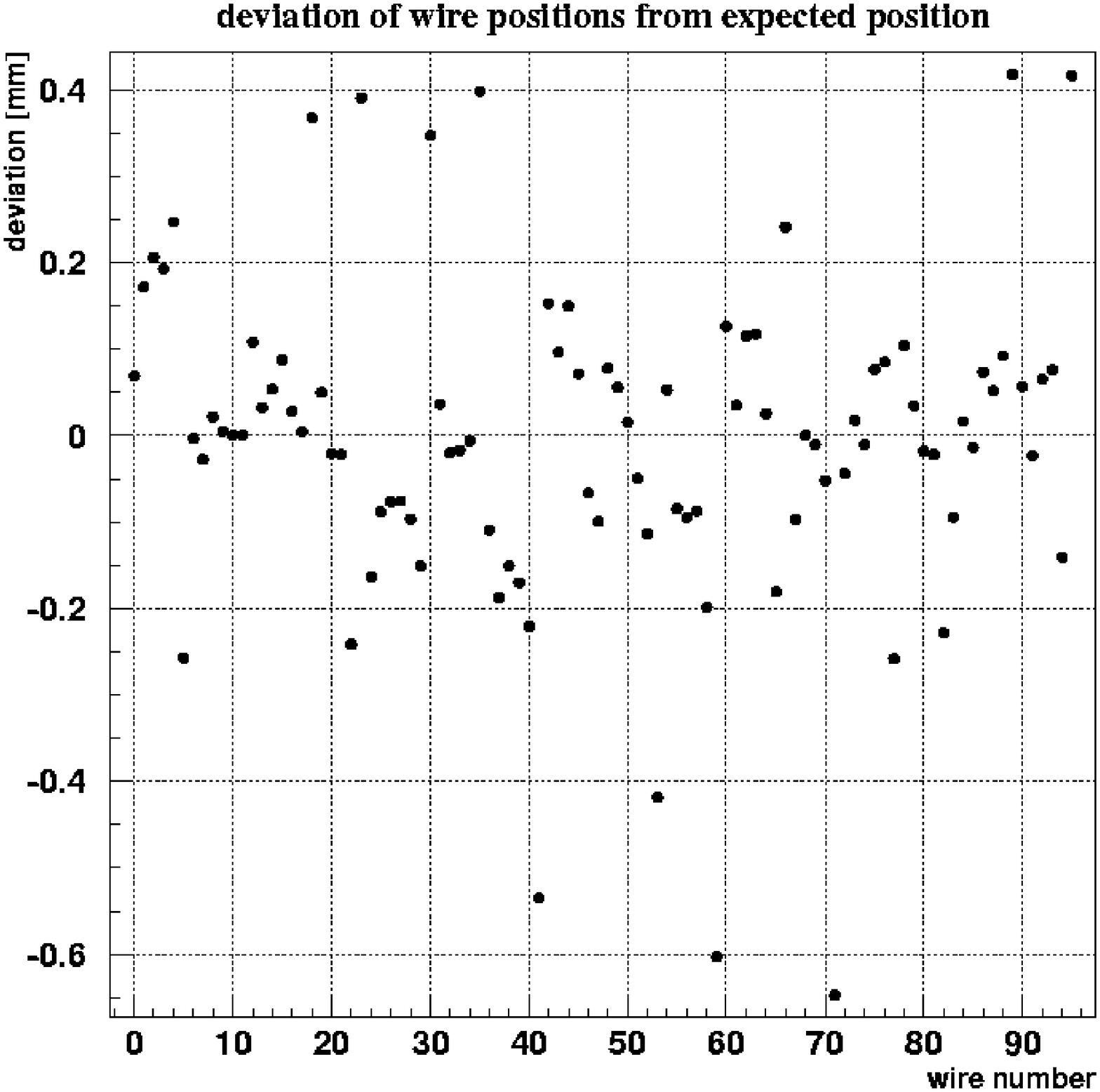}
    \includegraphics[scale=0.23,clip]{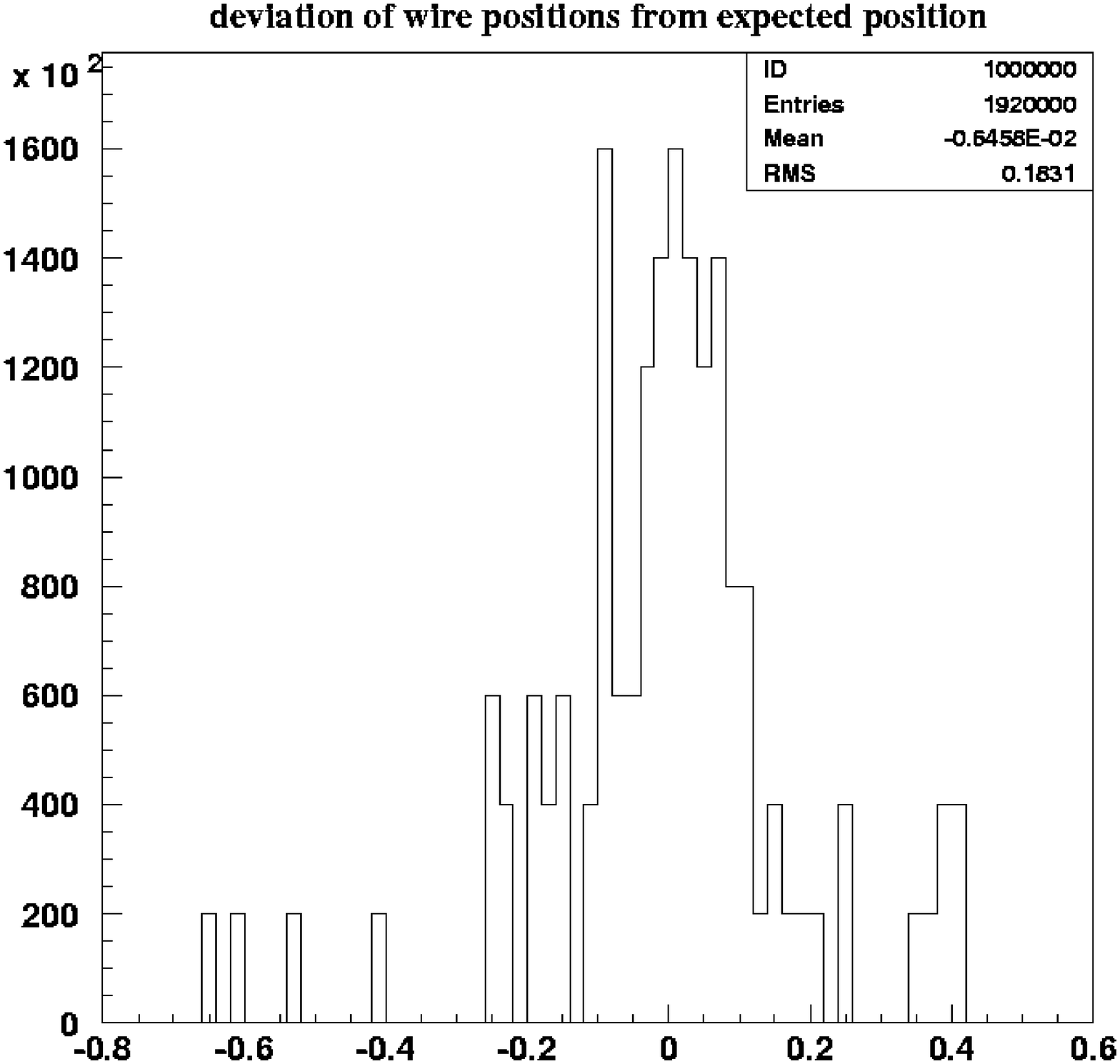}
    \includegraphics[scale=0.23,clip]{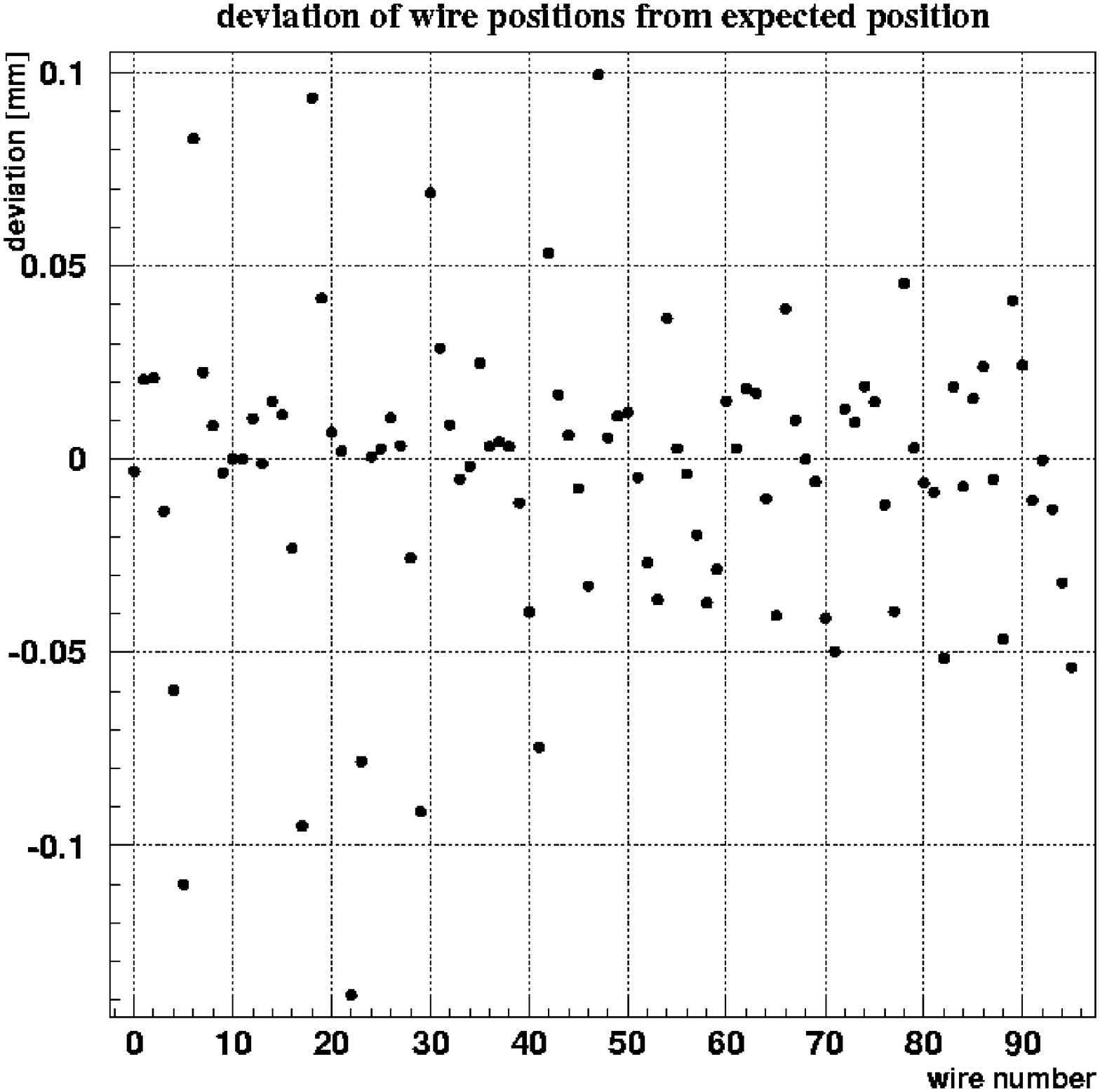}
    \includegraphics[scale=0.23,clip]{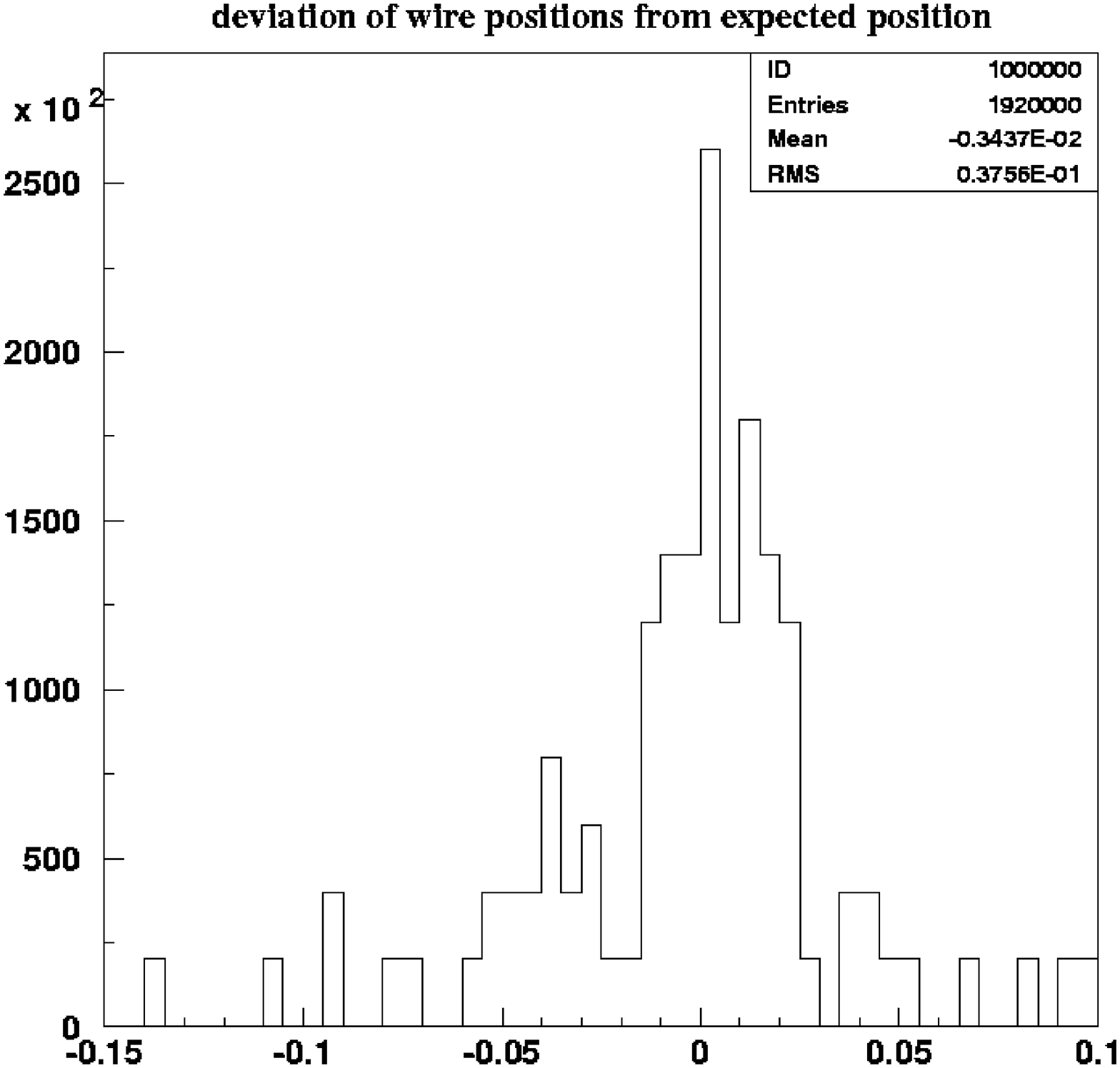}
  \end{center}
  \caption[Wire position before and after the wire alignment]{Wire position before (upper plots) and after the wire alignment (lower plots). In the left plots the deviation of the wire position is shown, in the right plots the distribution of the deviation.\label{wirecal}}
\end{figure}

In most cases, e.g. for a narrow angle distribution, it is more sufficient to align only one wire coordinate (perpendicular to the track axis). A simple way is an iterative procedure using the mean value of the residual distribution. For each iteration the wire position will be corrected by the deviation of this mean value from zero, attenuated by a scale factor of 0.3 to 0.5. It is of advantage that not only tracks perpendicular to the wire deviation have to be used. Tracks at angles around those also give good results and act as additional attenuation. This method is very stable, but without a scale factor the technique is not working and the wire correction starts to oscillate. Furthermore this method only works for homogeneously irradiated drift tubes, for tubes at the edges the position correction is biased and should not be used. Fig. \ref{wirecal} shows some plots of the wire positions before (upper plots) and after the correction (lower plots). In the left plots the deviation of the wire position is shown, in the right plots the distribution of the deviation can be seen. The RMS value gives the overall mean wire misalignment. The plotted results show the improvement of the accuracy of the wire position from 183~$\mu$m down to 38~$\mu$m after five iterations.
\subsection[Calibration schematic]{Calibration schematic}
\begin{figure}[htbp]
  \begin{center}
    \includegraphics[scale=1.15,clip]{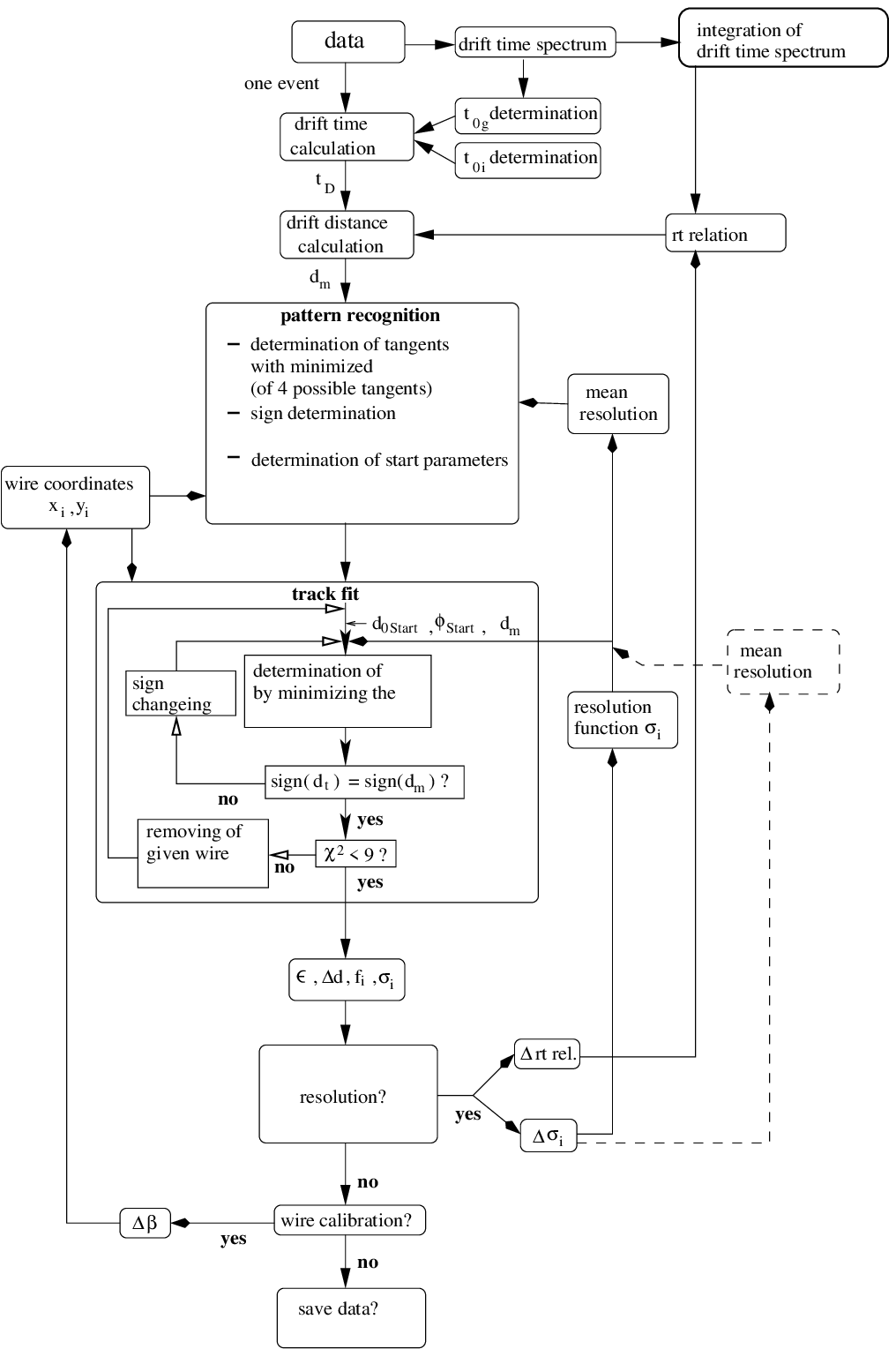}
  \end{center}
  \caption[Schematic of calibration]{Schematic of the calibration. For the first iterations (dashed lines) a mean resolution is used. The wire calibration is done once for the determination of the wire positions.\label{picpap}}
\end{figure}

A flow diagram of the iterative calibration procedure to determine the $rt$--relation and the resolution function is presented in Fig. \ref{picpap}. For the first iteration a $rt$--relation from the integration of the TDC spectrum is used. Furthermore, the run time correction ($t_0$), not described in this article, can be extracted from the TDC spectra for each channel and used for the drift time calculations. For the first iteration a constant resolution function $\langle \sigma \rangle = 1000~\mu$m is used in the pattern recognition and the track fit. The procedures of the pattern recognition and the track fit are described in detail in chapter \ref{sec:pat} and \ref{sec:trackFit}. After the iteration the resulting $rt$--relation is used as start parameter for the next iteration. This procedure will be repeated until the $rt$--relation will not change. The indicator is the sum of the squared differences of old and new $rt$--relation. Now the  procedure described above will be repeated using the resolution function. The convergation of the $rt$--relation and the resolution function appears after several iterations (typically 10). At the end a wire calibration can be done if necessary. In this case, the whole calibration has to be repeated, starting from the drift distance calculation (see Fig. \ref{picpap}). After several iterations (typically 5), the wires are aligned. 

\end{document}